\documentstyle[twocolumn,aps,psfig,epsf]{revtex}

\def\EQ{\begin{equation}}
\def\EN{\end{equation}}
\def\EQA{\begin{eqnarray}}
\def\ENA{\end{eqnarray}}
\begin{document}
\title{Scaling laws and vortex profiles in 2D decaying turbulence}
\author{Jean-Philippe Laval$^{1}$,
Pierre-Henri Chavanis$^{2}$, B\'ereng\`ere Dubrulle$^{1,3}$
and Cl\'ement Sire$^{2}$} 
\address{
$^{1}$ 
CEA/DAPNIA/SAp L'Orme des Merisiers, 709, F-91191 Gif sur Yvette, France\\
$^2$  Laboratoire de Physique Quantique (UMR5626 du CNRS), Universit\'e
Paul Sabatier, F-31062 Toulouse Cecex 4, France\\
$^3$ Observatoire Midi-Pyr\'en\'ees (UMR 5572 du CNRS), 14, av. E.
Belin, F-31400 Toulouse, France}

\maketitle

\begin{abstract} 
We use high resolution numerical simulations over several hundred of
turnover times to study the influence of small scale dissipation onto
vortex statistics in 2D decaying turbulence. A self-similar scaling
regime is detected when the scaling laws are expressed in units of
mean vorticity and integral scale, as predicted in \cite{CMPWY91},
and it is observed that viscous effects spoil this scaling
regime. This scaling regime shows some trends toward that of the
Kirchhoff model, for which a recent theory predicts a decay exponent
$\xi=1$ \cite{Sire}. In terms of scaled variables, the vortices have a
similar profile close to a Fermi-Dirac distribution.
\end{abstract}


\vspace{0.1cm}

\narrowtext

In recent years, two-dimensional turbulence has received rather large
interest because of its applications in astrophysics and geophysics
and its relative accessibility to numerical simulations with respect
to fully developed three-dimensional turbulence. Two-dimensional flows
are characterized by the presence of coherent structures (vortices)
which play the role of elementary particles and dominate the
dynamics. More precisely, the relaxation of 2D decaying turbulence is a
three-stage process: during an initial transient period, the fluid
self-organizes from random fluctuations and a population of coherent
vortices emerges. Then, when two like-sign vortices come into contact
they merge and form a bigger structure.  As time goes on, the vortex
number decreases and their average size increases, in a process
reminiscent of a coarsening stage. Finally, when only one dipole is
left, it decays diffusively due to inherent viscosity.

Two types of studies have been conducted to characterize this
relaxation process: some have focused on the precise structure of the
vortices (vorticity profiles, $\omega-\psi$ relationships...)  while
others described how the average vortex properties (typical radius,
core vorticity, vortex number...) evolve with time. There has been some
attempts to predict the final state (the dipole) in terms of
statistical mechanics of the 2D Euler equation as developed in
\cite{MRS}.  It is found that a prediction from the initial condition
leads to incorrect results due to the effect of viscosity which
dissipates the high order moments of the vorticity during the long
evolution of the flow towards that state.  However, if the constants
of the motion are evaluated at later times (i.e. before the last
merging) the prediction gets better and better \cite{BP}. This implies
that the statistical theory cannot predict the final state of a long
viscous evolution but is likely to describe correctly the structure of
a vortex that forms after a rapid merging. It was therefore suggested
that the isolated vortices of 2D turbulence are sort of
quasi-equilibrium states or ``maximum entropy bubbles''
\cite{bubbles}. In this respect, 2D vortices share some common
features with stellar systems like elliptical galaxies which also
undergo a mixing process during a phase of ``violent relaxation''
\cite{galaxies}.

Other authors have chosen to disregard the precise structure of the
vortices in order to study how the typical characteristics of the flow
evolves with time. In such experiments and numerical simulations, it
is found that the vortex density $n$, their average radius $a$ and
their typical core vorticity $\omega$ seem to follow power laws.  Two
different scenarii have been proposed. In the first one, known as the
Batchelor theory \cite{Batch69}, the assumption of a unique invariant
(the energy $E\sim n\omega^2 a^4$), and the occurrence of a unique
relevant time scale $\omega^{-1}\sim t$ implies that the number of
vortices $n$ decays as $n\sim t^{-\xi}$, with $\xi=2$. This theory
also implies the occurrence of a unique length scale, the typical
distance between vortices $n^{-1/2}$, of the same order of magnitude
as their typical radius $a$. Hence, the total area occupied by the
vortices $na^2$, or alternatively the Kurtosis $K\sim (n a^{2})^{-1}$,
remains constant, implying $a\sim t$, while the enstrophy $Z\sim
n\omega^2 a^2$ decays like $\sim t^{-2}$. This is consistent with the
selective decay hypothesis, which states that for slightly viscous
flows, the enstrophy is dissipated while the energy remains
essentially conserved.

However, hyperviscous Navier-Stokes simulations \cite{mcwilliams90a}
and experiments \cite{hansen98} suggest a different scenario in which
the typical core vorticity $\omega$ is an additional
invariant. Assuming $n\sim t^{-\xi}$ and the energy conservation such
that $na^4$ is now constant, the scaling theory consistent with this
scenario \cite{CMPWY91} leads to the slow decrease of the total area
occupied by the vortices $na^2\sim t^{-\xi/2}$ (or $a\sim
t^{\xi/4}$). The enstrophy now decays as $Z\sim t^{-\xi/2}$ while the
Kurtosis increases as $K\sim t^{\xi/2}$. The occurrence of an extra
dimensionless relevant parameter $na^2$ prevents the determination of
$\xi$ from purely dimensional grounds as was done within Batchelor
theory.

From the numerical and experimental side, the situation is rather
confused.  Matthaeus {\it et al.} \cite{MSMOM91} performed a very long
Direct Numerical Simulation of the Navier-Stokes equation (DNS) and
found that the enstrophy decays approximatively like $t^{-1}$.  More
recently, other DNS at very large resolution \cite{Chas97} produced a
similar decay rate $Z \sim t^{-0.8}$. By contrast, in numerical
simulations using hyperviscosity (HDNS) \cite{weiss93} the enstrophy
decays like $Z\sim t^{-0.3}$. These hyperviscous simulations show an
overall agreement with the second scenario with an exponent $\xi\sim
0.75$.

In a first series of experiments for which 3D effects were not fully
controlled, Tabeling and collaborators \cite{cardoso} obtained scaling
laws compatible with the first scenario (roughly conserved vortex area
coverage), but with $\xi\approx 0.44$ instead of $\xi=2$. In a second
series of experiments with stratification \cite{hansen98}, the same
group obtained scaling laws in favor of the second scenario with
$\xi\approx 0.7$. In both case, dissipation is provided not via a
standard viscosity, but mainly via friction at the bottom of the
experimental apparatus. A simple rescaling however can make the
experimental system equivalent to a real 2D system, with a time
dependent viscosity \cite{hansen98}.

Theoretical attempts have been made to understand and clarify the
decay process. Among them, simple models describing vortex aggregation
process have used point vortices following a Kirchhoff-Hamilton
dynamics and merging via empirical rules derived from imposed
conservation laws. The model corresponding to the second scenario with
constant energy and core vorticity was first investigated in
\cite{weiss93}, leading to $\xi\sim 0.75$ \cite{weiss93,sirevor}, in
agreement with HDNS and experiments. Recently, the same model was
investigated using a renormalization group procedure which allows for
much larger simulation times (3 more decades in time) \cite{Sire}.
The true scaling regime is only obtained for times much larger than
previous simulation or experimental times, and the asymptotic decay
exponent is found to be $\xi\approx 1$. Interestingly, in the time
range comparable to that of HDNS and experiments, the function $n(t)$
displays a pseudo-scaling range with an effective exponent $\xi\simeq
0.7$. An effective three-body theory shows that the decay of the total
area occupied by vortices results in a situation where mergings occur
principally via three-body collisions.  A kinetic theory based on
these three-body processes leads to $\xi=1$ in agreement with the
simulations. However, since the conservation laws are built in the
model {\it a priori}, there is a definite need for more precise
comparisons with DNS.

Motivated by this observation, we have undertaken numerical simulations
of $2D$ turbulence at high resolution $2048^2$, using both normal and
hyperviscosity. The goal of these simulations was two-fold: first,
determine which of the two scenarii is more appropriate to describe
the decline, and whether there is an influence of the numerical scheme
used to dissipate energy; second, determine whether there is an
asymptotic transition between the value $\xi\simeq 0.7$ usually reported,
and a value $\xi\approx 1$ predicted by the Kirchhoff model, or any
other value. This imposes to consider a large number of initial
vortices, so that the decay of their number occurs over several
decades of time.\\

Both viscous and hyperviscous simulations were performed with a
pseudo-spectral code with periodic boundary conditions. We chose the
resolution so that the typical size of initial vortices was outside
the dissipative range. We used a $2048^2$ grid for viscous simulations
and a $1024^2$ grid for hyperviscous ones.  A random vorticity field
was introduced as initial conditions.  The energy spectrum
corresponding to this initial field is given by $E(k) =
{k^{30}}/{(k+k_0)^{60}}$.  Most of the energy is concentrated at the
wavenumber $k_0 = 100$.  This corresponds to a situation with
approximately 10000 vortices randomly localized.  The identification
of the vortices is based on the vortex selection procedure defined by
McWilliams \cite{mcwilliams90a}. We used the same procedure but with
a new definition of the vorticity threshold based on both the
vorticity maximum of the total field and the vorticity maximum of the
given vortex.\\

\begin{figure}
%      \vspace{5cm}
 \epsfxsize=8.8cm\epsfbox{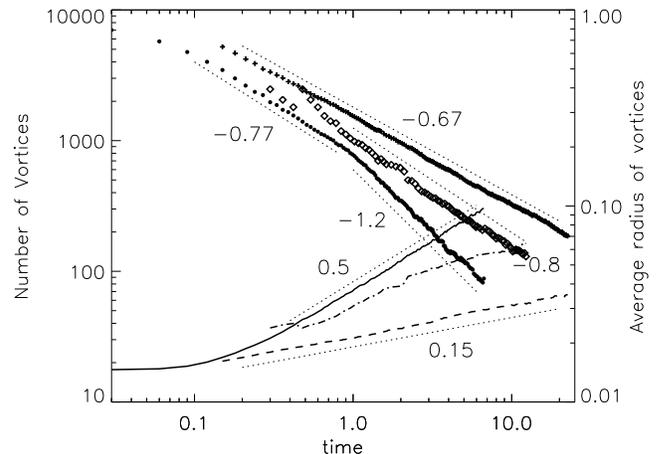} \caption[]{ Evolution of
the number of vortices and their average radius for the three
simulations ($\bullet$ and $-$: DNS; $+$ and $-\ -\ $: HDNS;
$\diamond$ and $- \cdot -$: subgrid scale model). }
         \label{fig_fig1}
   \end{figure}

A good summary of the different scaling laws detected in our
simulation is provided by Fig. \ref{fig_fig1}. The results of the
simulation with hyperviscosity mostly confirm the previous numerical
simulations performed over shorter time scales. The number of vortices
decays like $n\sim t^{-0.67}$ over two decades in time, while the
average vortex radius and the Kurtosis increase like $a\sim t^{0.15}$
and $K\sim t^{0.30}$ respectively. The enstrophy decays steadily over
the simulation like $Z\sim t^{-0.40}$, while the energy remains almost
constant, especially at the later stages. Finally, the maximum of
vorticity is almost conserved, decaying approximatively like
$t^{-0.12}$ in a first stage, even slower (like $t^{-0.06}$) in the
late stage of the simulation. These last scalings are only
approximate, since we did observe strong local increase of the maximum
of vorticity, which we associate with strong steepening of the local
vorticity profile within a few isolated vortices. This is probably an
artifact of the hyperviscosity, as was previously reported. Overall,
these results are compatible with the second scenario with $\xi\approx
0.7$.\

When normal diffusion is adopted instead of hyperdiffusion, the
behavior changes dramatically. Two different regimes can be clearly
distinguished: in the first one, between $t=0$ and $t\approx 1$ (or
equivalently until the total number of vortices has decayed by one
order of magnitude), a clean power law $n\sim t^{-0.77}$ can be
observed for the total number of vortices, which is close to that
obtained with hyperviscous computations. There is also a rather clean
scaling law $Z\sim t^{-1.3}$ for the enstrophy within vortices
decaying much steeply than in the hyperviscous computations.  For any
other quantities, a monotonic but non scaling behavior is observed,
with a decrease of the vorticity maximum and of the energy, and a slow
increase of the average radius and of the Kurtosis.  In the second
regime ($t\gtrsim 1$), rather clean scaling laws for most quantities
suddenly emerge, and become markedly different from the corresponding
hyperviscous ones. The number of vortices decays like $n\sim t^{-1.2}$
and the average radius increases like $a\sim t^{0.50}$ resulting in an
almost constant vortex area coverage and Kurtosis. This regime cannot
however be described by the Batchelor theory since $\xi\simeq 1.2$
instead of $\xi=2$ and, in addition, $\omega\sim t^{-0.6}$ and $Z\sim
t^{-1.3}$.

To test if the observed discrepancy is an effect of the finite
viscosity, we have also performed inviscid computations using a
turbulent subgrid scale model described in \cite{LDN00}. Two different
scaling regimes are observed: in the early stage of the simulation the
Kurtosis remains almost constant like in the Batchelor theory but the
vortex density decays with an exponent $\xi\sim 0.9$ instead of
$\xi=2$. At later times, the Kurtosis increases like $K\sim t^{0.3}$,
$\omega$ becomes nearly constant, and we observe scaling laws
compatible with the second scenario with $\xi\sim 0.8$. During all the
course of the simulation, the energy is nearly constant, like in the
hyperviscous case.\

\begin{figure}
%      \vspace{5cm}
\epsfxsize=8.8cm\epsfbox{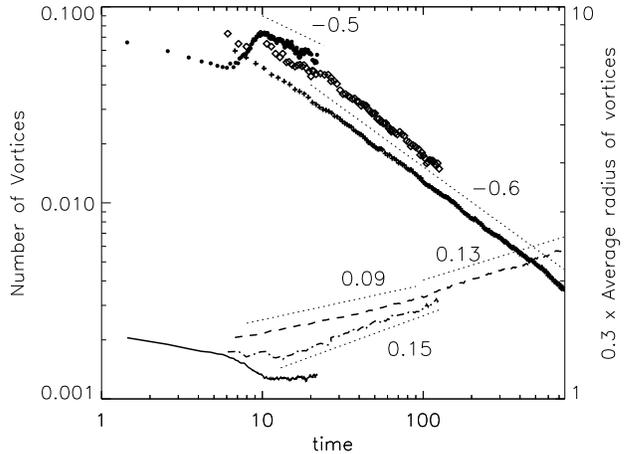}\caption[]{Same as
Fig. \ref{fig_fig1} when the number of vortices, their radius and the
time have been rescaled appropriately. }
         \label{fig_fig2}
\end{figure}

Since all these results were obtained using the same initial
conditions, they show that the vortex statistics is strongly
influenced by the dissipative process acting at small scale. In
particular, the ``absolute'' scaling law exponents, obtained when the
quantities are normalized with large scale quantities such as the
initial energy and the size of the box, are not universal. In the
spirit of the scaling theory, it may however be interesting to
determine whether a universality class can be detected with
appropriate {\sl local} rescaling of the quantities. The most obvious
choice is to use as a unit of time the inverse ``average'' vortex
vorticity (assumed to be constant in the standard scaling theory)
defined as $\bar \omega=\sqrt{KZ}$ where $K$ and $Z$ are the kurtosis
and enstrophy of the vortices. As a unit of length, it is then natural
to consider the integral scale $\lambda=\bar\omega/\sqrt{E}$ built
with $\bar\omega$ and the energy $E$. These local units are in fact
implicitly taken in the Kirchhoff model.  The scaling laws obtained
with these units are reported on Fig. \ref{fig_fig2}. We now obtain a
much better agreement between the hyperviscous and the inviscid
computations, where the scaling laws seem to be compatible with the
scaling scenario with $\xi\sim 0.6$, while the two regimes of the
viscous simulation seem to collapse, apart from a small transition
zone, into a single regime where $\omega/\bar\omega\sim (\bar \omega
t)^0$, $\lambda^2 n(t)\sim (\bar\omega t)^{-0.6}$, $a/\lambda\sim
(\bar \omega t)^{0.15} $ and $Z/(\bar \omega)^2\sim K^{-1}\sim (\bar
\omega t)^{-0.33}$.

\begin{figure}
%      \vspace{5cm}
\epsfxsize=8.8cm\epsfbox{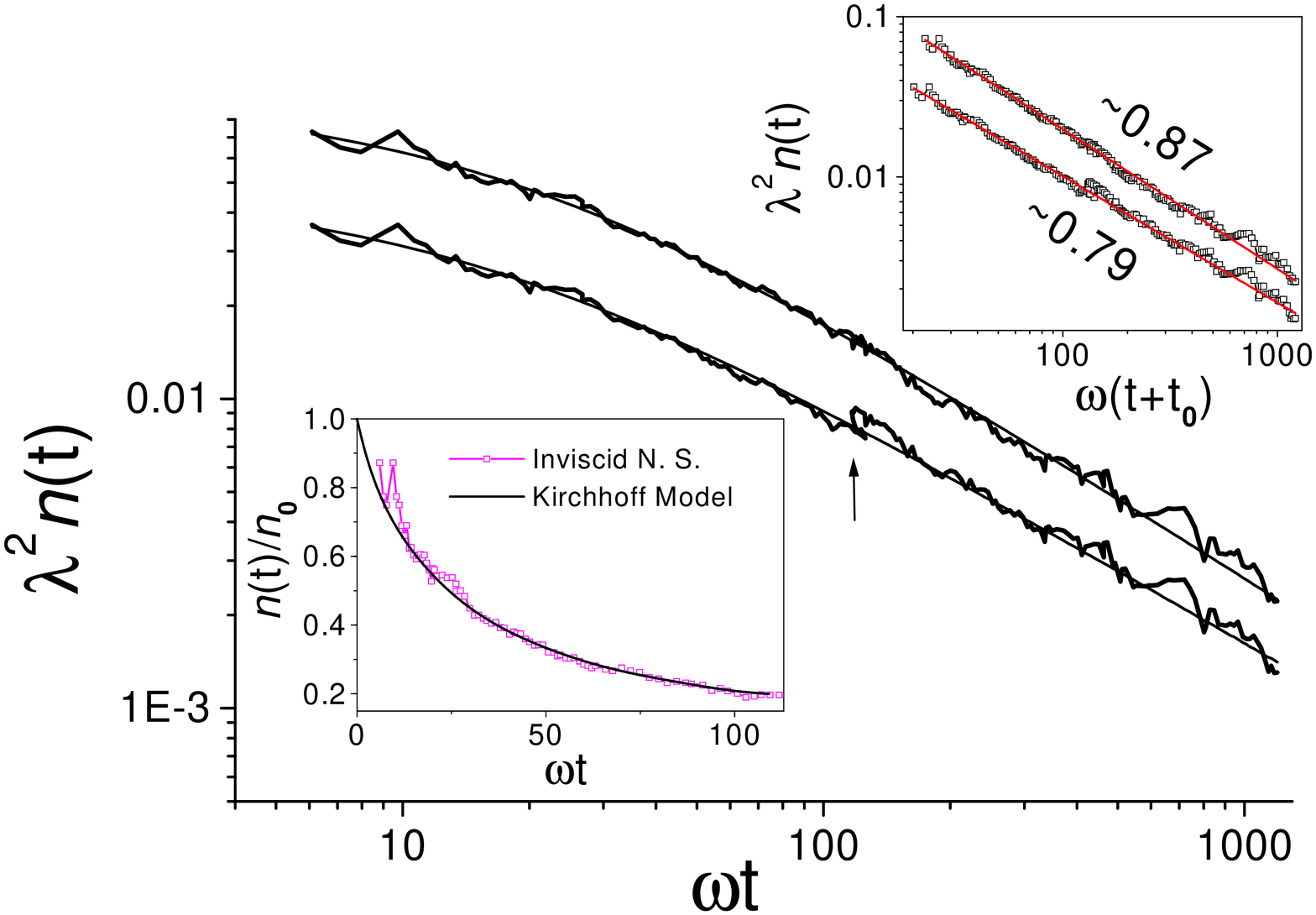} \caption[]{Long-time
evolution of the rescaled vortex density obtained using inviscid
simulations. In the top curve the discontinuity arising from a change
of subgrid (arrow) has been smoothed out. The best fits to the
functional form $n(t)=n_0/(1+t/t_0)^\xi$ are displayed. The same data
are plotted as a function of $t+t_0$ in the top inset. The effective
exponents obtained show some trends towards the asymptotic Kirchhoff
model value $\xi\approx 1$ \cite{Sire}. For short times, coinciding
with times before the change of subgrid, the data are in agreement
with the best available Kirchhoff model {\it polydisperse} simulations
\cite{weiss93,sirevor} (bottom inset).}  \label{fig_fig3} \end{figure}

The scaling laws obtained in these units are now reminiscent of the
early stage of the Kirchhoff simulation. To test whether these scaling
laws steepen into a regime in which $\xi=1$, one needs to continue the
simulations over one or two more decades in time, which would
represent several months of continuous integration using our numerical
resources. This makes longer integrations of the viscous or
hyperviscous case impossible. However, in the inviscid case, we tried
to use the flexibility of the subgrid scale model to move the large
scale/small scale cut-off towards larger scales (following the
behavior of the integral scale), thereby allowing a gain of
computational time from 10 to 100. We started the simulation with the
vorticity field from the DNS at $t=0.3$ when the energy is small
enough at the higher wavenumber. The use of a coarser grid induces a
small loss of enstrophy and Kurtosis via the filtering procedure used
to change the cut-off scale in our model. We observed that this change
of cut-off produces an artificial, cut-off dependent, new scaling
regime in``absolute" scaling coordinates (using the length of the box
and the energy as units), but a universal scaling regime in ``local"
scaling coordinates (using the integral scale and the mean vorticity
as units). This universal regime is shown in Fig. \ref{fig_fig3} for
$n(t)$ and clearly suggests a local exponent $\xi$ effectively
increasing with time. The extrapolated value for $\xi$ is $\xi\sim
0.79$ and $\xi\sim 0.87$ when the artificial discontinuity due to the
change in resolution has been smoothed out (by multiplying the curve
after the subgrid change by a factor $\sim 0.86$).\

Our results therefore support the validity of a universal self-similar
evolution of the vortices for inviscid, or nearly inviscid decaying
turbulence. This self-similar scenario appears universal when
appropriate local units are considered, and a single exponent in the
range $\xi=0.8\sim 0.9$ is found, compatible with that of the
Kirchhoff model. Finally, viscous effects tend to favor the
conservation of the vortex coverage area $n a^{2}$ and modify the
scaling exponents without, however, leading to the Batchelor model.

\begin{figure}
%      \vspace{5cm}
 \epsfxsize=8.8cm\epsfbox{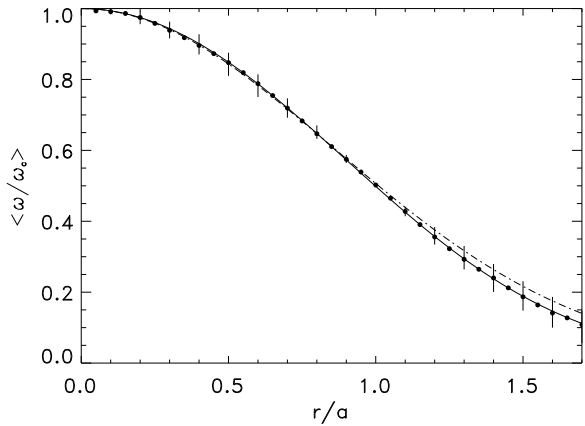} 
\caption[]{Average vortex profile fitted by a Fermi-Dirac distribution
(full line) and a Gaussian profile (dash-dot).}
\label{fig_fig4}
\end{figure}

We have also found that the vortices present a universal profile when
the vorticity is normalized by the central vorticity and the distance
by the typical vortex radius defined by the condition
$\omega(a)=\omega(0)/2$. This profile is represented on
Fig. \ref{fig_fig4} and has been obtained by averaging over $\sim 30$
vortices at different times in the inviscid simulation. The error bars
indicate to which extent this profile can be considered as
``universal''. As time goes on, these bars become smaller showing a
trend towards a self-similar evolution. We observe that the
Fermi-Dirac distribution $\omega=\sigma_{0}/(1+\lambda e^{\alpha
r^{2}})$ provides a very good fit to this profile while the Gaussian
distribution is less accurate (but has only one fitting parameter). In
the statistical theory of 2D turbulence, the Fermi-Dirac distribution
maximizes the mixing entropy introduced by \cite{MRS} at fixed
circulation and angular momentum. Interestingly, it corresponds to the
stationary solution of the equation obtained by \cite{quasilinear}
from a quasilinear theory of the 2D Euler equation.

\end{document}